\documentclass[twocolumn,showpacs,preprintnumbers,amsmath,amssymb]{revtex4}

\usepackage{graphicx}
\usepackage{dcolumn}
\usepackage{bm}

\newcommand{\beq}{\begin{equation}}
\newcommand{\eeq}{\end{equation}}    

\newcommand{\bea}{\begin{eqnarray}}
\newcommand{\eea}{\end{eqnarray}}    

\newcommand{\beqn}{\begin{eqnarray}}
\newcommand{\eeqn}{\end{eqnarray}}

\def\beq{\begin{equation}}
\def\eeq{\end{equation}}

\begin{document}

\preprint{...}


\title{The interaction of gravitational waves with strongly magnetized plasmas
}

\author{Heinz Isliker}
 \email{isliker@helios.astro.auth.gr}
\author{Loukas Vlahos}%
 \email{vlahos@helios.astro.auth.gr}
\affiliation{%
Section of Astrophysics, Astronomy and Mechanics,\\
Department of Physics, University of Thessaloniki,\\
GR 54006 Thessaloniki, GREECE
}%



\date{\today}

\begin{abstract}
We study the interaction of a gravitational wave (GW) 
with a plasma that is strongly magnetized.
The GW is considered a small disturbance, and the plasma
is modeled by the general relativistic analogue of the induction equation 
of ideal MHD and the single fluid equations.
The equations are derived without neglecting any of the non-linear 
interaction terms, and 
the non-linear equations are integrated numerically.
We find that 
for strong magnetic fields of the order of $10^{15}\,$G
the GW excites electromagnetic plasma waves very close to the 
magnetosonic mode. The magnetic and electric field oscillations 
have very high amplitude, 
and a 
large amount of energy is absorbed from the GW by 
the electromagnetic 
oscillations, 
of the order of $10^{23}\,$erg/cm$^3$ in the case presented here. 
The absorbed energy is proportional to $B_0^2$, with $B_0$ the background
magnetic field.
The energization of the plasma takes place on fast time scales 
of the order of milliseconds.
The amount of absorbed 
energy is comparable to the energies emitted 
in the most energetic astrophysical events, such as 
giant flares on magnetars and possibly even short Gamma ray bursts 
(GRB),
for which the mechanism analyzed here also has the fast  
time-scales required.
\end{abstract}

\pacs{04.30.Nk, 04.25.Dm, 52.35.Mw, 95.30.Sf, 95.30.Qd}
\keywords{gravitational waves --- plasma physics}
\maketitle


Gravitational waves (GW) can carry a large amount of energy
near the sources where they are generated (e.g.\ \cite{Kokk04}).
They tend not to interact much with matter under normal conditions,
it has been shown though
in a number of articles
(e.g.\ \cite{Ignatev95},
\cite{Brod99}, \cite{Brod00}, 
\cite{Serv00b}, \cite{Papadop01},
\cite{Serv03}, 
\cite{Moor03}, 
\cite{Kall04}
\cite{Moor04}) that 
GWs excite various kinds of plasma waves, the more efficient, the 
stronger the background magnetic field is and the more tenuous the plasma is. 
Most of these studies are analytical
and the equations describing the GW-plasma interaction 
were linearized, only 
Ref.\ \cite{Ignatev95} made
an analytical study of a non-linear model,
Refs.\ \cite{Brod99} and \cite{Kall04} took
some second order effects into account, 
and Ref.\ \cite{Duez05} 
performed a numerical study.
The GW-plasma interaction is 
a totally non-linear 
effect, and there is so-far no conclusive answer to the question 
of how much energy can be absorbed by a plasma from a GW. 

Here, we study the GW-plasma interaction 
in its
full non-linearity,  solving the non-linear system of equations numerically.
Our main interest is in the amount of energy absorbed by the plasma
from the GW and in the kind of plasma waves excited by the GW, and
we focus on the case of very strong magnetic fields of the order of
$10^{15}\,$G.

The interaction of gravitational waves (GWs) with plasmas, if it is 
efficient in transferring energy from the GW to the plasma,
might well be the basic mechanism 
behind the most energetic astrophysical events, such as  
giant flares on magnetars (i.e.\ highly energetic 
outbursts in the stars' magnetospheres; see e.g.\ \cite{Stella05}) 
or possibly even short Gamma ray bursts (GRB; see e.g.\ \cite{Piran04};
magnetars are strongly magnetized neutron stars
[see e.g.\ \cite{Duncan92}], which at the same time 
appear as Soft Gamma Ray repeaters [SGR]).
The origin of short GRBs is far less established
than that of the long GRBs. One model recently discussed is that 
short GRB might be giant flares on 
magnetars. 
In this model, the flares are caused by a catastrophic 
reconfiguration
of the stellar magnetic field (e.g.\ \cite{Hurley05}, \cite{Nakar05}, 
\cite{Schwartz05}).
The energy released in short GRBs is currently estimated to be at least
$10^{48}\,$erg \cite{Nakar05}, the uncertainty is mainly due to 
the problem of associating short GRBs with physical objects.
The duration of short GRBs is less than 2 seconds. Giant flares
have similar durations and release energies in the range 
$10^{44}\,$ to $10^{46}\,$erg (e.g.\ \cite{Stella05}).

Our results 
show that a GW generated by a magnetar might well
be the source of the energy released in a giant flare
and may-be even in a short GRB, through the 
absorption of the GW energy by a 
plasma in the vicinity
of the magnetar. In this scenario, the GW-plasma interaction is the 
primary mechanism that energizes the plasma before 
a giant flare or may-be a short GRB.   
Giant flares and possibly short GRBs would then be an indirect 
observation of GWs.


The GW is considered as a small amplitude perturbation of 
the otherwise flat spacetime, and we assume it to be 
$+$ polarized and to 
propagate along the $z$-direction, so that the metric has 
the form
$
g_{ab} = {\rm diag}(-1,1+h,1-h,1)
\label{metric}
$, 
with $h(z,t)<<1$ the amplitude of the GW \cite{Hartle03}. 
Our aim is to express the final equations in terms of 
the potentially observable 
quantities (electric field $\vec E$, magnetic field $\vec B$,
3-velocity $\vec V$ of the fluid, and rest-mass density $\rho$), 
which can either be defined in a Local Inertial Frame (LIF)
or in an orthonormal frame (ONF) \cite{Hartle03}. 
Here, we use the ONF, because it is a global frame, and it can  
be shown that the ONF in our case is locally equivalent to a LIF 
when applying the particular coordinate transformation given in 
\cite{Landau81}. The ONF has the advantage that 
4-vectors and tensors take the same form as in flat space-time,
the effects of curvature appear only through the covariant 
derivatives,
which are calculated 
with use of the
Ricci rotation coefficients \cite{Landau81}.
Indices of quantities
in the ONF carry a hat in the following.

We assume an ideal conducting fluid, 
so that the electric field is given by the ideal 
Ohm's law, $0=F^{\hat{a}\hat{b}}u_{\hat{b}}/c$,
which in the ONF takes the usual form,
$
0 = \hat{\gamma} \left(\vec E + \frac{1}{c} \vec V \times \vec B\right) 
$,
with $\hat{\gamma} = 1/\sqrt{1-\vec V^2/c^2}$,
$F^{\hat{a}\hat{b}}$ Faraday's field tensor,
$u^{\hat{a}}$ the 
4-velocity, $u^{\hat{a}}=\hat{\gamma}(c,V_x,V_y,V_z)$,
and $c$ the speed of light.
The evolution of the magnetic field is determined by 
the Maxwell's equation, 
$
F_{\hat{a}\hat{b};\hat{c}} + F_{\hat{b}\hat{c};\hat{a}} 
            + F_{\hat{c}\hat{a};\hat{b}} = 0 
\label{farad}
$
\cite{Landau81}.
The electromagnetic (EM) energy momentum tensor is defined as 
$
T^{\hat{a}\hat{b}}_{(EM)} =  
   \frac{c^2}{4\pi} \left(
  F^{\hat{a}\hat{c}}F^{\hat{b}}_{\ \hat{c}} 
               -\frac{1}{4}\eta^{\hat{a}\hat{b}} 
                 F^{\hat{c}\hat{d}}F_{\hat{c}\hat{d}}\right)
$,
and for the fluid, we have the energy momentum tensor 
$
T^{\hat{a}\hat{b}}_{(fl)} = H u^{\hat{a}} u^{\hat{b}} + 
\eta^{\hat{a}\hat{b}} p c^2
\label{Tflm}
$,
where $H$ is the enthalpy, $p$ the pressure, and 
$\eta_{\hat{a}\hat{b}}=\text{diag}(-1,1,1,1)$ the metric of flat 
spacetime \cite{Landau81}. We 
assume an ideal and adiabatic fluid, so that 
$
H = \rho c^2 + \frac{p}{\Gamma-1} + p
\label{enthm}
$,
with $\Gamma$ the adiabatic index.
The total energy momentum tensor 
$T^{\hat{a}\hat{b}} = T^{\hat{a}\hat{b}}_{(fl)} + T^{\hat{a}\hat{b}}_{(EM)}$
yields
the momentum and energy equations 
$
T^{\hat{a}\hat{b}}_{\ \ ;\hat{b}} = 0 
$
\cite{Landau81}.
Continuity is expressed by 
$\left(\rho u^{\hat{a}}\right)_{;\hat{a}} = 0 .
\label{cont}$
The evolution of the GW is determined by the linearized Einstein equation,
where we take the back-reaction of the plasma onto the GW into account
\beq
-\partial_{tt} h + c^2\vec\nabla^2 h 
           = -\frac{1}{2}\frac{16\pi G}{c^4}\left(\delta T_{xx}-\delta T_{yy}\right) ,
\label{gw}
\eeq
where $\delta T_{xx}$, $\delta T_{yy}$ are the non-background, fluctuating 
parts of the components $T_{xx}$, $T_{yy}$ of the total energy momentum tensor,
and $G$ is the gravitational constant \cite{Hartle03}.
To close the system of equations, we assume an adiabatic and isentropic 
equation of state,
$
p = K \rho^{\Gamma}
\label{eos}
$,
with $K$ a constant.


We focus on the excitation of MHD modes which
propagate in the $z$-direction, parallel to the propagation
direction of the GW and perpendicular to the background magnetic
field $\vec{B_0}=B_0 { \bf e}_{\hat{x}}$. We let consequently 
$\vec B\Vert { \bf e}_{\hat{x}}$
$\vec E\Vert { \bf e}_{\hat{y}}$ and 
$\vec V\Vert { \bf e}_{\hat{z}}$, and all variables depend spatially only on 
$z$ (note that $B_x$ in the following is the total magnetic field, it 
includes $B_0$). 
In specifying the general equations to this particular
geometry, (i) we express all 4-vector and tensor components
through the potentially observable $B_x$, $V_z$, and $\rho$;
(ii) we expand the covariant derivatives; (iii) we keep all non-linear 
terms, no approximations are thus made (except for the linearized 
Einstein equation). In this way, we are led to a system of 
non-linear, coupled, partial differential equations in a spatially 
1-D geometry:
With the electric field $E_y$ from Ohm's law,  
$
E_y  =  - \frac{1}{c}V_z B_x  
$,
Faraday's equation is fully expanded to 
\beq
\partial_t B_x = c \partial_z E_y +\frac{1}{2} B_x \frac{\partial_t h}{1-h}
-\frac{1}{2} c E_y \frac{\partial_z h}{1-h} .
  \label{farady}
\eeq
Expansion of the $z$-component of the momentum equation
yields
\bea
\partial_t q_{\hat{z}} + \partial_z
        \left[\left(q_{\hat{z}} - 
                            \frac{c}{4\pi}\left(-E_y B_x\right)\right)V_z 
              \right]  \nonumber\\
        + \partial_z
        \left[\frac{c^2}{8\pi}\left(B_x^2+E_y^2\right)\right]  
        + c^2\partial_z p\nonumber\\
        +\frac{c}{8\pi}B_x\left(cB_x + E_y V_z\right) 
                  \frac{\partial_z h}{(1+h)} \nonumber\\
        -\frac{c}{8\pi}E_y\left(cE_y+B_xV_z\right) 
                    \frac{\partial_z h}{(1-h)} \nonumber\\
        -q_{\hat{z}}\left(\partial_t h +V_z\partial_z h\right) \frac{h}{1-h^2}
= 0  ,
\label{momz}
\eea
where we defined the new momentum variable $q_{\hat{z}}$ as
$
q_{\hat{z}}:=
                    H V_z\hat{\gamma}^2 
           + \frac{c}{4\pi}\left(-E_y B_x\right)
$.
The continuity equation takes the form
\beq
\partial_t D + \partial_z (D V_z) 
        - \left(D \partial_t h +D V_z\partial_z h\right)
                \frac{h}{1-h^2}
                 = 0  ,
\label{contz}
\eeq
with the new density variable $D:=\hat{\gamma}\rho$.
The GW evolves according to 
\beq
\partial_{tt} h 
=  c^2\partial_{zz} h 
             + \frac{2 G}{c^2}
              \left(E_y^2-(B_x^2-B_0^2)\right) 
             + \frac{16\pi G}{c^2} (p-p_0) h  ,
\label{gwz}
\eeq
where the background magnetic field $B_0$ and background pressure 
$p_0= K \rho_0^{\Gamma}$ have
been subtracted.

To recover $\hat{\gamma}$, $\rho$, $p$, $H$, and $V_z$ from 
the explicitly evolving variables
$q_{\hat{z}}$, 
$D$, and $B_y$, we solve the definition of $q_{\hat{z}}$ for $V_z$ and 
insert it into a reformulated definition of $\hat{\gamma}$, which 
yields
a non-linear equation for
$\hat{\gamma}$, which we solve numerically. Once
$\hat{\gamma}$ is recovered, all the other primary variables follow
in a straightforward way.


We solve the GW-plasma system of equations applying 
a pseudo-spectral method that is based on Chebyshev polynomials
(see e.g.\ \cite{Forn98}). 
Time stepping is done with the method of lines, using a fourth order
Runge-Kutta method with adaptive step-size control.
The one-dimensional grid along the $z$-direction 
consists of 256 grid-points and corresponds to a physical domain 
along the $z$-axis of length $L=5.4\,10^7\,$cm.
The sampling time step $\Delta t$ is set to 
$\Delta t=T_{gw}/14$, with $T_{gw}=1/f_{gw}$ and $f_{gw}$ the 
GW frequency.


We assume a background magnetic field $B_0$ of  
$10^{15}\,$G, a background density
$\rho_0 =10^{-14}\,$g cm$^{-3}$, and
an adiabatic index $\Gamma = 4/3$. 
The initial conditions are $B_x(z,0)=B_0$, 
$V_z(z,0)=0$, $\rho(z,0)=\rho_0$, and $h(z,0)=0$.
The GW has as boundary condition at the left end $z_L$ of the box
$h(z_L,t) = h_0(t) \cos(k_{gw}z_L-\omega_{gw}t)$,
so that a monochromatic plane wave is entering the box.
The amplitude $h_0$ rises within roughly $1\,$ms 
from $0$ to $10^{-4}$, at which value it stays constant for about $3\,$ms, 
where after it decays to $0$ again. 
At the right edge $z_R$ of the box, we apply non-reflecting 
boundary conditions to $h(z_R,t)$.
$B_x$, $V_z$, and $\rho$ have free outflow boundary conditions at both 
edges of the box.
The GW frequency is set to $f_{gw}=5\,$kHz, and the GW dispersion relation 
is of the form $2\pi f_{gw}=\omega_{gw}=k_{gw}c$, with $k_{gw}$ the wave-number
of the GW.


\begin{figure}
\resizebox{6truecm}{!}{\includegraphics{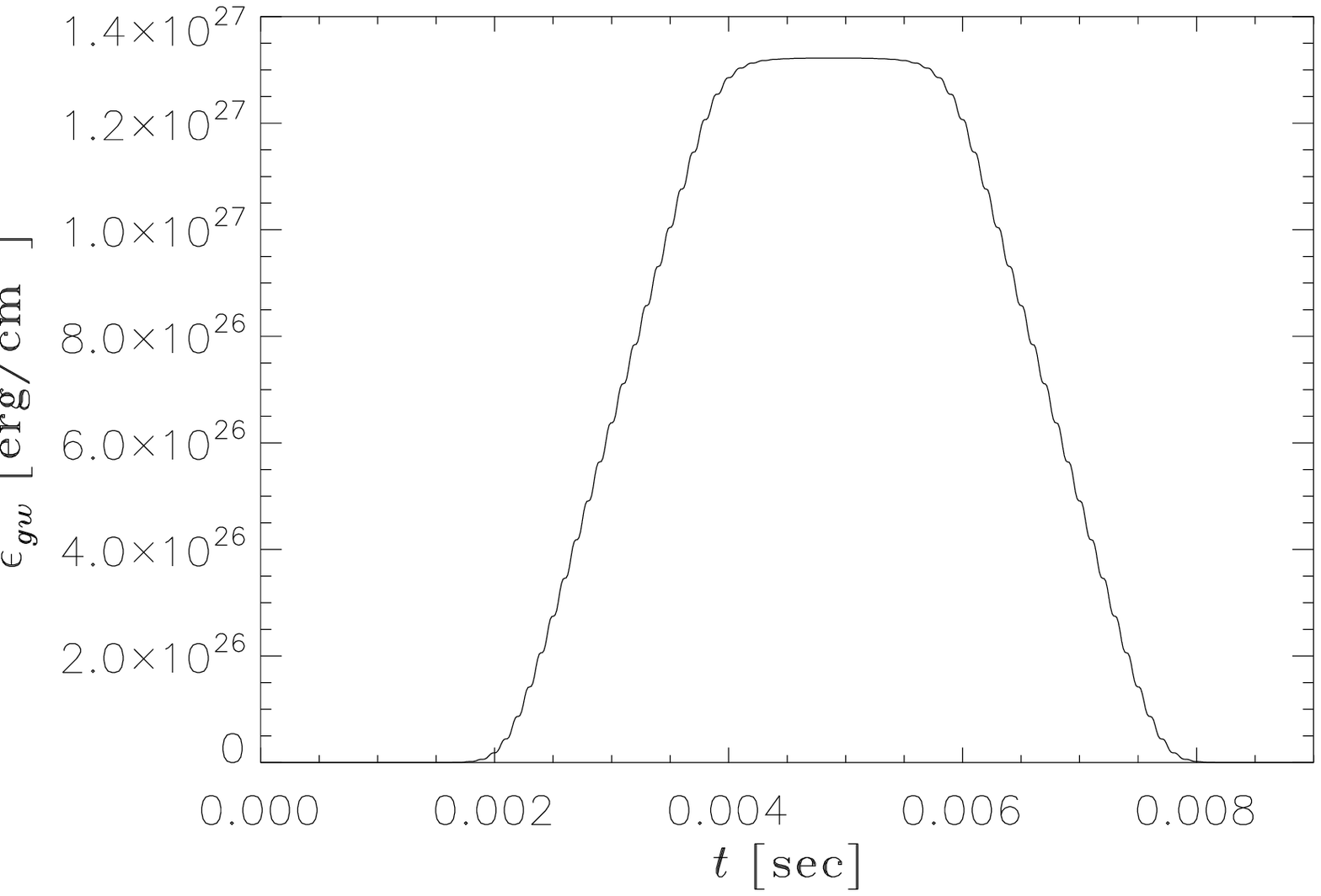}}
\resizebox{6truecm}{!}{\includegraphics{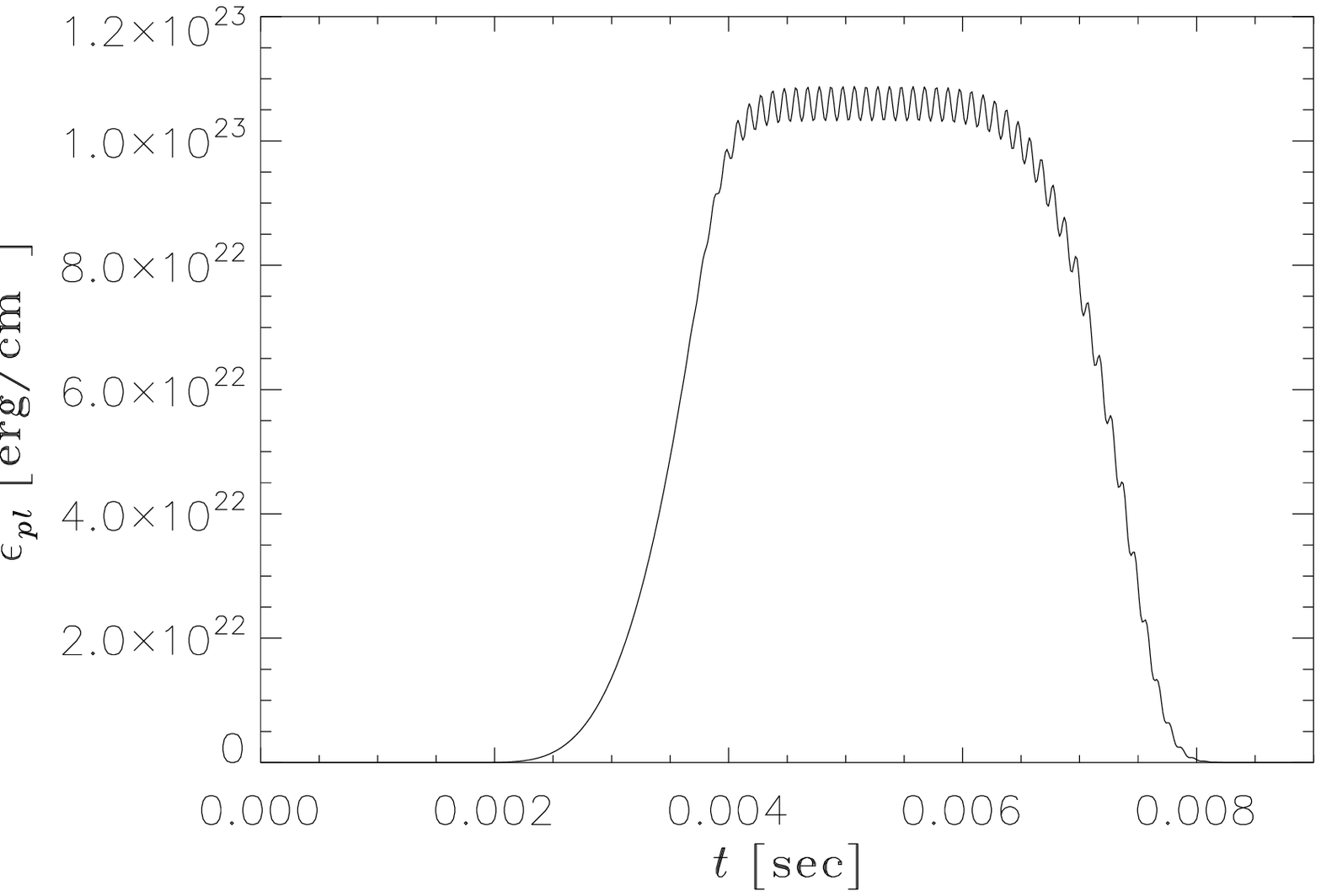}}
\caption{
Top: Average energy density $\epsilon_{gw}(t)$ of the GW as a function of time.
--- Bottom: Mean total energy density $\epsilon_{pl}(t)$ of the plasma as a function of time.
\label{reltotalenergy}
}
\end{figure}

The total mean energy density $\epsilon_{pl}$ in the plasma 
at a given time $t$ is numerically determined as
\begin{eqnarray}
\epsilon_{pl}(t) &=&
\left[\frac{1}{8\pi} \int E_y(z,t)^2\,dz
+\frac{1}{8\pi} \int (B_x(z,t) - B_0)^2\,dz \right. \nonumber \\
&& \left. 
+ \frac{1}{2} \int \rho(z,t) v_z(z,t)^2\,dz \right] / L
\end{eqnarray}
with $L$ the size of the system
 --- note that we subtract 
the constant background magnetic field
$B_0$ in order to take into account only the energy that is
in the wave motion. 
Fig.\ \ref{reltotalenergy} shows $\epsilon_{pl}(t)$ 
and the mean energy density $\epsilon_{gw}(t)$ of the GW 
as a function of time, where
$\epsilon_{gw}(t) = \frac{c^2}{32\pi G} \omega_{gw}^2 \bar{h}(t)^2$,
with $\bar{h}(t)$ the mean instantaneous amplitude of the GW oscillation,
defined as the root mean square average over the entire simulation box
\cite{Hartle03}.
At maximum GW amplitude, the energy density of the 
GW amounts to $\epsilon_{gw} = 1.3\times 10^{27}\,$erg/cm$^3$.
Once the GW enters the system, the plasma starts to absorb
energy from the GW, and in roughly $2\,$ms, slightly delayed in the beginning
but finally in parallel with the GW 
reaching its maximum energy,
the absorption has reached its maximum, the energy density in the 
plasma is roughly $10^{23}\,$erg/cm$^3$. 
The absorbed energy is a fraction $10^{-4}$
of the GW energy density, so that the back-reaction onto the GW
is not yet important.  
When the GW leaves the system,
the energy in the plasma decays almost together with the GW amplitude,
i.e.\ the excited waves propagate out of the simulation box.

The GW 
excites wave motions in the 
plasma that travel with the GW and
whose amplitudes increase linearly towards the out-flowing edge of the 
box. Maximum amplitudes attained are 
$3\times 10^{12}\,$statvolt/cm ($9\times 10^{16}\,$V/m)
for the electric field, 
$3\times 10^{12}\,$G for 
the oscillating 
magnetic field, 
roughly three
orders of magnitude less than $B_0$, and finally the fluid velocity oscillations
have a maximum amplitude $8\times 10^7\,$cm/sec. 
The fluid motions thus remain non-relativistic, so that our non-relativistic
estimate of the kinetic energy is justified.



\begin{figure}
\resizebox{6truecm}{!}{\includegraphics{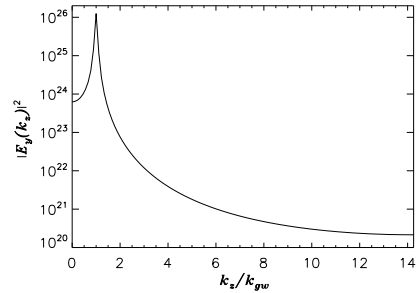}}
\caption{
Spatial Fourier transform $\vert\hat{E}_y(k_z,t)\vert^2$ of $E_y(z,t)$ at a fixed 
time $t=0.00545\,$s. 
\label{spectra}
}
\end{figure}

The waves excited in $E_y$, $B_x$, and $V_z$ have wave-number 
$k_z$ and frequency $\omega$ that cannot be distinguished from
$k_{gw}$ and $\omega_{gw}$ within the numerical precision of the
simulation, see Fig.\ \ref{spectra}. In particular, we do not 
find any harmonics to be  excited. 
The relativistic Alfv\'en speed $u_A^2=B_0^2/(4\pi\rho_0+B_0^2/c^2)$
is very close to the speed of light ($(c-u_A)/c=2\times 10^{-24}$) so that 
the excited plasma modes are indistinguishable from magneto-sonic modes.

The numerical results show that, for strong magnetic fields, 
a large amount of  
energy is absorbed by the plasma from the GW on a short time-scale, 
which is of the
order of 10 GW periods, i.e.\ in the millisecond range. 
The energy
absorbed by the plasma is thus not proportional to the 
duration of the GW-plasma interaction if this duration is longer 
than typically a few GW periods.

\begin{figure}
\resizebox{6truecm}{!}{\includegraphics{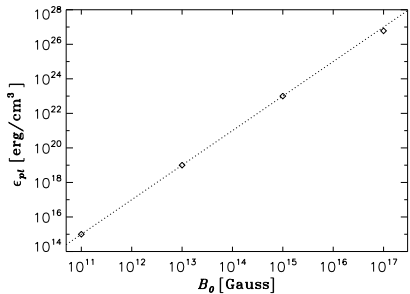}}
\caption{
The energy density $\epsilon_{pl}$ absorbed by the plasma 
(at maximum absorption) as a function of the 
background magnetic field $B_0$ (diamonds), together with a reference line of 
logarithmic slope 2 (dotted). 
\label{EnB0}
}
\end{figure}

\begin{figure}
\resizebox{6truecm}{!}{\includegraphics{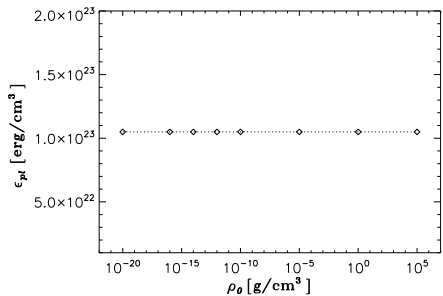}}
\caption{
The energy density $\epsilon_{pl}$ absorbed by the plasma 
(at maximum absorption) as a function of the 
background matter density $\rho_0$ (diamonds, 
connected with a dotted line). 
\label{Enrho0}
}
\end{figure}

In a parametric study, we found that the energy absorbed by the plasma
is proportional to $B_0^2$ (see Fig.\ \ref{EnB0}) and to $h_0^2$. 
Varying $\rho_0$ in the range 
$10^{-20}\,{\rm g/cm^3}\leq \rho_0\leq 10^5\,{\rm g/cm^3}$,
it turned out that the absorbed energy is independent of
the value of the matter density $\rho_0$, see Fig.\ \ref{Enrho0}, 
the kinetic energy is actually negligible compared to the electromagnetic
energy (this is in accordance with the fact that the background matter 
rest-energy density is much smaller than the background magnetic 
energy density in the entire range of values $\rho_0$ investigated).
The absorbed energy density is furthermore proportional to $\omega_{gw}^2$,
as we verified by varying the frequency in the kHz range 
(from $1$ to $10\,$kHz).
It also seems that the time for the plasma
needed to reach the maximum level of energy absorption is related to the
time needed for the GW to cross the box, $L/c$, which equals
$0.002\,$sec for the case considered here.

The maximum amplitudes of the excited oscillations are proportional to the
box length $L$, as explained above.
This
implies that the amount of energy density absorbed is proportional to the 
squared box-length, $L^2$,
with the physical meaning of $L$ to be the length along the propagation 
direction of the GW where the GW meets a 
constant magnetic field. 
We additionally verified this scaling behavior with numerical 
simulations in which 
the box-length $L$ was varied. 
The absorbed energy is thus proportional to $L^3$ 
and to the effective area $A_{eff}$ 
through which the GW is incident on the plasma, where 
$A_{eff}:=V/L$, with $V$ the volume in which the interaction takes place.
We can summarize our numerical findings for the total energy $E_{pl}$ 
absorbed by the
plasma as follows, noting that in the case presented here 
the absorbed energy through an effective area of $1\,$cm$^2$ is $L\times 10^{23}\,$erg, 
with $L=5.4\times 10^7\,$cm, 
\begin{eqnarray}
E_{pl}&=&3.4\times 10^{7} \left(\frac{L}{1\,{\rm cm}}\right)^3  
       \left(\frac{A_{eff}}{1\, \text{cm}^2}\right) 
       \left(\frac{B}{10^{15}\,{\rm G}}\right)^2 \nonumber\\
  &&     \ \ \ \ \ \ \ \ \times \left(\frac{h_0}{10^{-4}}\right)^2
                           \left(\frac{f_{gw}}{5\,\text{kHz}}\right)^2
         \,\text{erg},
\label{energy}
\end{eqnarray}
and $E_{pl}$ is independent of $\rho_0$.

{\it Astrophysical Application.}
%
%
On the basis of our results, we can suggest a new model for the 
primary mechanism behind giant flares on magnetars and possibly even 
behind short GRBs.
This mechanism
deposits a large amount of energy 
in a plasma
on time-scales of milli-seconds,
prior to the actual outburst of the giant flare or possibly the short GRB. 
In our model, we assume that
a magnetar generates a GW, which
travels away from the magnetar and
interacts with a plasma in the vicinity of the magnetar, whereby 
the plasma efficiently absorbs energy from the GW. 
Once the plasma is energized, secondary mechanisms will be triggered that 
convert the energy ultimately to Gamma rays, the nature of these 
mechanisms is though not in the scope of this article.

The typical magnetic field $B_\star$ of a magnetar is currently 
estimated to be 
of the order of $10^{15}\,$ to $10^{16}\,$G at the surface
(e.g.\ \cite{Nakar05}, \cite{Hurley05}, \cite{Stella05}).
GWs are expected to be emitted by magnetars, 
as by usual neutron stars, 
when they 
undergo some deformation, e.g.\ 
as the result of
the rearrangement of the strong stellar magnetic field. 
In such a case, the GW amplitude at 5 stellar 
radii typically is $10^{-4}$ (e.g. \cite{Kokk04}).
At the same distance,
the background spacetime can be considered flat.
The magnetic field is poloidal within the light cylinder,
so that, near the equatorial plane, 
a GW generated by the magnetar and traveling radially outwards 
propagates perpendicular to the magnetic field,
as in the presented 
simulations.

\begin{figure}
\resizebox{6truecm}{!}{\includegraphics{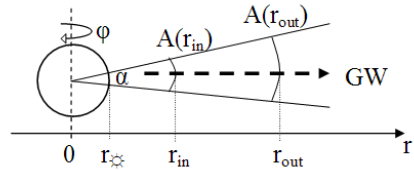}}
\caption{
Sketch: The GW interacts with a plasma volume in the equatorial 
plane (see text for details).
\label{sketch}
}
\end{figure}

\begin{figure}
\resizebox{6truecm}{!}{\includegraphics{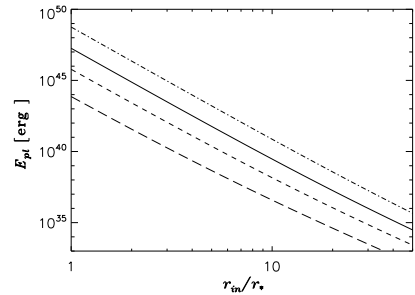}}
\caption{
Estimate of 
the energy $E_{pl}$ absorbed by the plasma 
as a function
of the inner distance $r_{in}$ from the star, for different interaction 
lengths $L$ 
(long dashes: $L=100\,$km, 
short dashes: $L=250\,$km, 
solid: $L=500\,$km, and dash-dotted: $L=1\,000\,$km),
and for $\alpha=10^o$, $\varphi=180^o$. 
\label{en_est}
}
\end{figure}

In order to get a rough estimate of the energy absorbed by a plasma,
we consider the plasma in the conical volume of radial size $L$ between the 
inner and outer radii $r_{in}$ and $r_{out}=r_{in}+L$, respectively 
(with $r$ the distance from the magnetar), 
which is limited to within the poloidal opening angle $\alpha$
and the toroidal range $\varphi$, as illustrated by the sketch in 
Fig.\ \ref{sketch}. 
We assume the magnetic 
field and the GW amplitude to be constant over the length $L$, 
with values $B_0=B_\star(r_\star/r_{in})^3$ and $h_0=h_\star(r_\star/r_{in})$,
respectively, in order 
to be able to apply 
Eq.\ (\ref{energy}) 
(note that the magnitude of the magnetic field along the radial direction 
is proportional to $1/r^3$). The plasma volume is given as  
$V(r_{in},L)
=\frac{2}{3}\,\varphi\sin(\alpha/2)\left(r_{out}^3-r_{in}^3\right)$, 
with corresponding effective area $A_{eff}=V(r_{in},L)/L$.

For a numerical estimate, we set  
$B_\star=10^{16}\,$G, $h_\star=10^{-4}$, $r_\star=10\,$km,
$\alpha=10^o$, and, to take possible anisotropic GW emission 
into account, $\varphi=180^o$.
Fig.\ \ref{en_est} shows
the absorbed energy as a function of $r_{in}$ for four different values
of $L$.
The energy decreases fast with distance, it roughly is proportional 
to $1/r_{in}^8$,
so that at $r_{in}=5\,r_\star$, the energy has fallen to  
$E_{pl}=10^{43}\,$erg for e.g.\ $L=500\,$km.
Extrapolating our results to regions closer to the star,
we  find at $r_{in}=2\,r_\star$ and again for $L=500\,$km a plasma energy 
$E_{pl}=10^{45}\,$erg, which is of the 
order of the energy released in giant flares on magnetars
(e.g.\ \cite{Nakar05}, \cite{Hurley05}, \cite{Stella05}). 
Even closer
to the star, at $r_{in}=r_\star$, the energy is of the order 
of $E_{pl}=10^{47}\,$erg, well in excess of the energy in giant flares,
and approaching the energy observed in short GRBs
according to latest 
estimates (e.g.\ \cite{Piran04}, \cite{Nakar05}). 
For different values of $B_\star$ and $h_\star$, 
the energy values given here scale according to Eq.\ (\ref{energy})
in a straightforward way.


Again for $L=500\,$km, the volume is $2\times 10^{7}\,$km$^3$ for 
$r_{in}=r_\star$ and increases to $4\times 10^{7}\,$km$^3$ for
$r_{in}=10\,r_\star$, 
which corresponds to effective areas of $A_{eff}=5\times 10^{4}\,$km$^2$ and
$A_{eff}=8\times 10^{4}\,$km$^2$, respectively. The involved volumes and 
areas are thus relatively small.


Our numerical simulations were done for the case of a flat background 
spacetime, which does not hold in the range $r_{in}\leq 2 r_\star$ anymore, 
to which we have extrapolated the energy estimates. 
The non-flatness of spacetime close to the star, together with   
the necessarily arbitrary choice of a value for $L$, imply 
that our energy estimates must be interpreted with some care,
they can though be taken indicative of the fact that the 
GW-plasma interaction is efficient and is 
an important mechanism near the star. 
Also to mention is the 
uncertainty concerning the actual
magnetic topology in the flaring magnetosphere,
since any deviation
from the dipole is likely to intensify the GW-plasma interaction 
through the enlargening of the possible interaction regions. 
It thus remains to be seen in how far the given numbers will be modified
when a more realistic decay of the background magnetic field 
is used in the 
simulations and when the curvature of the background spacetime is included, 
which though only can be done when extending the 
equations to the case of three spatial dimensions. 
In favor of the model of giant flares on magnetars (and, less certain, but
possibly also of short GRBs) driven by GWs is that the mechanism proposed
has a fast enough time-scale, of the order of milli-seconds.


{\it In summary,} our results show that strongly magnetized 
plasmas are 
efficient absorbers of GWs,
largely irrespective of the plasma density, and with an absorption
time-scale of the order of milli-seconds. 
This implies that GWs may be the energy source
for secondary, highly energetic phenomena. It also implies that GWs 
may eventually be strongly damped, 
if appropriate conditions are met. In particular, we can conclude that 
the GW-plasma interaction is an efficient and important 
mechanism in magnetar atmospheres close to the star.

\begin{acknowledgments}
This work was supported by the Greek Ministry of Education through the
PYTHAGORAS program. We thank K.\ Kokkotas, D.\ Papadopoulos, N.\ 
Stergioulas, and J.\ Ventura for helpful discussions.
\end{acknowledgments}


\end{document}